\documentclass[aps,prl,twocolumn,showpacs,preprintnumbers,superscriptaddress,floatfix,amsmath,amssymb,amsfonts]{revtex4-1}
\pdfoutput=1
\usepackage{graphicx}
\usepackage{dcolumn}
\usepackage[bookmarks=true,colorlinks,linkcolor=blue,urlcolor=blue,citecolor=blue]{hyperref}
\usepackage{color}
\usepackage{subfigure}
\usepackage{amsmath, amssymb}
\usepackage{verbatim}
\usepackage{bm}
\usepackage{mathrsfs}

\providecommand{\up}{\uparrow}
\providecommand{\dn}{\downarrow}

\makeatletter
\@tfor\next:=abcdefghijklmnopqrstuvwxyzABCDEFGHIJKLMNOPQRSTUVWXYZ\do{%
  \def\command@factory#1{%
    \expandafter\def\csname bf#1\endcsname{\mathbf{#1}}
  }
 \expandafter\command@factory\next
}
\@tfor\next:=abcdefghijklmnopqrstuvwxyzABCDEFGHIJKLMNOPQRSTUVWXYZ\do{%
  \def\command@factory#1{%
    \expandafter\def\csname cal#1\endcsname{\mathcal{#1}}
  }
 \expandafter\command@factory\next
}
\makeatother

\newcommand{\ket}[1]{\left|#1\right>}

\newcommand{\braket}[1]{\left<#1\right>}
\newcommand{\para}[1]{\left(#1\right)}
\newcommand{\abs}[1]{\left|#1\right|}

\newcommand{\COMMENT}[1]{}

\begin{document}

\title{
Fibonacci Anyons From Abelian Bilayer Quantum Hall States}

\author{Abolhassan Vaezi}
\affiliation{Department of Physics, Cornell University, Ithaca NY 14850}
\author{Maissam Barkeshli}
\affiliation{Station Q, Microsoft Research, Santa Barbara, CA 93106-6105, USA}

\begin{abstract}
The possibility of realizing non-Abelian statistics and utilizing it for 
topological quantum computation (TQC) has generated widespread 
interest. However, the non-Abelian statistics that can be realized in most accessible proposals 
is not powerful enough for universal TQC. In this paper, we
consider a simple bilayer fractional quantum Hall (FQH) system with the $1/3$ Laughlin state
in each layer. We show that interlayer tunneling can drive
a transition to an exotic non-Abelian state that contains the famous `Fibonacci'
anyon, whose non-Abelian statistics is powerful enough for universal TQC. 
Our analysis rests on startling agreements from a variety of 
distinct methods, including thin torus limits, effective field theories, and coupled wire constructions. 
We provide evidence that the transition can be continuous, at which
point the charge gap remains open while the neutral gap 
closes. This raises the question of whether these exotic phases may 
have already been realized at $\nu = 2/3$ in bilayers, as past experiments
may not have definitively ruled them out. 
\end{abstract}
\maketitle

{\bf Introduction--} There is currently intense interest in the realization of exotic 
quantum phases of matter that host quasiparticles with non-Abelian statistics \cite{nayak2008,alicea2012review},
partially due to the possibility of topological quantum computation (TQC). 
While there are many promising candidate platforms for realizing non-Abelian statistics, 
almost all of them have the drawback that they are not powerful enough to realize universal TQC. 

Recently, it has been proposed that a wide class of non-Abelian
defects can be synthesized by starting with simple double-layer or single-layer fractional quantum Hall (FQH) 
states and properly including certain spatially non-uniform patterns of interlayer 
tunneling \cite{barkeshli2012a,barkeshli2013genon,barkeshli2013}
or superconductivity \cite{clarke2013,lindner2012,cheng2012,vaezi2013}. Subsequently, 
it was shown that by coupling these engineered non-Abelian defects in an appropriate 
manner, it is possible to realize exotic non-Abelian phases that
are powerful enough for universal TQC \cite{mong2013,vaezi2013b} \footnote{See also \cite{teo2011,ludwig2011,wires} for related work.}. 
However, engineering the interactions of these defects in a physically realistic setup is a major challenge. Nevertheless, these studies 
suggest the possibility that these exotic, computationally universal non-Abelian phases might 
be realized in a simpler fashion, by starting with either (1) conventional double layer FQH states (such as the $(330)$ state,
which contains independent  $1/3$ Laughlin states in each layer) and increasing the interlayer tunneling uniformly in space, 
or (2) conventional single layer FQH states, and uniformly increasing the coupling to a superconductor.

In this paper, we present two basic advances, mainly in the double layer context with interlayer tunneling. 
First, we show that the appearance of these computationally universal non-Abelian states can be 
understood in the thin torus limit, where the interlayer tunneling is taken to be uniform in space. 
In this limit we systematically derive the properties of the quasiparticles for large interlayer tunneling. 
These include the so-called `Fibonacci' quasiparticle, whose non-Abelian braiding statistics allow
for universal TQC. Second, we find the possibility of a continuous quantum phase 
transition between the conventional bilayer FQH states and these exotic non-Abelian ones, as the interlayer 
tunneling is increased. We show that this theory is described by a $SU(3)_1 \times SU(3)_1 \rightarrow SU(3)_2$ 
Chern-Simons-Higgs transition, and also provides a many-body wave function for the non-Abelian state.  
The startling agreement between these distinct approaches, and with the earlier constructions 
\cite{mong2013,vaezi2013b,vaezi2013c}, provides evidence that this non-Abelian state 
can be stabilized with uniform tunneling.  

\begin{figure}
\includegraphics[width=2.6in]{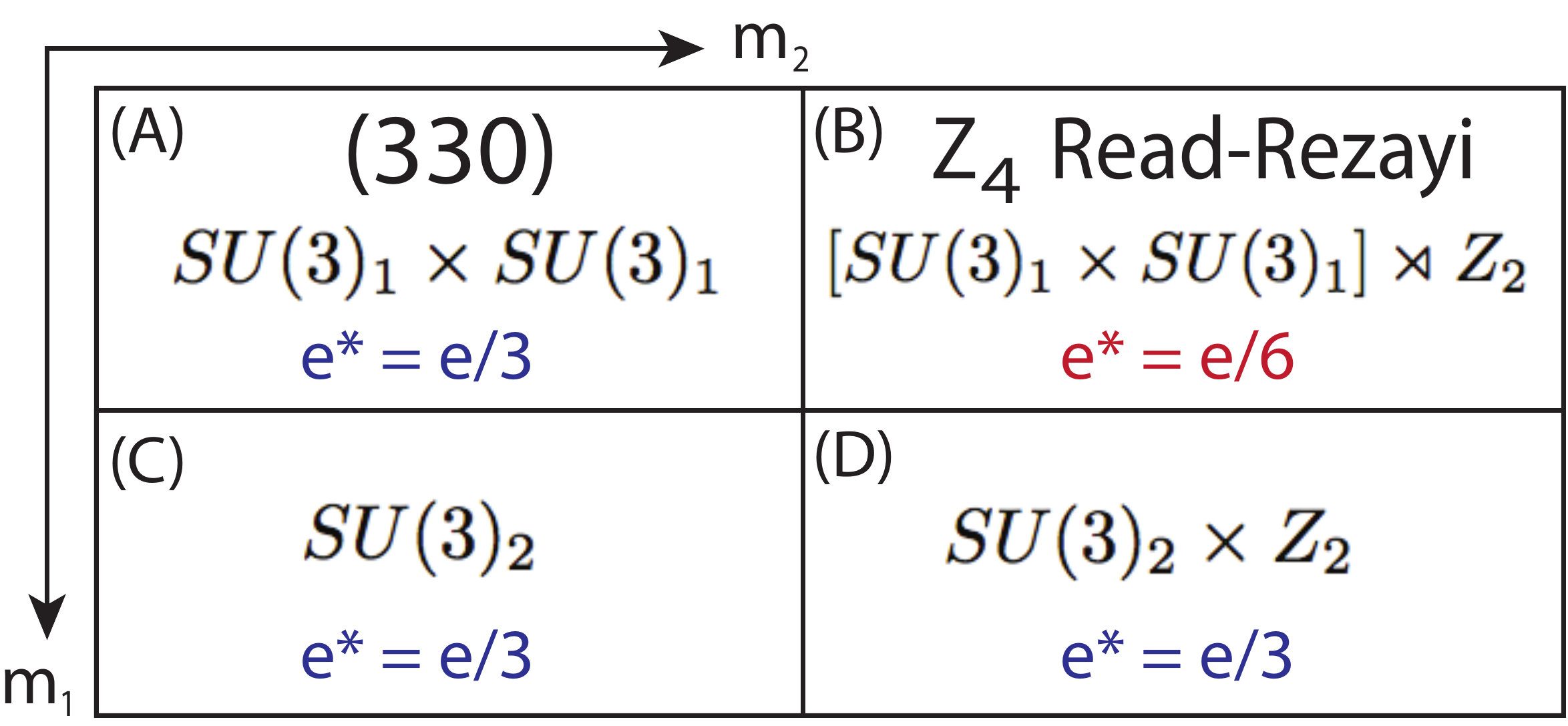} \caption{\label{pdFig} 
$\nu = 2/3$ proposed global phase diagram, for interlayer tunneling on the order of interaction strengths. 
We find possible continuous transitions between four different states, as described in the main text. 
(A) The $(330)$ state, (B) the $Z_4$ Read-Rezayi state,
(C) A non-abelian FQH state described by $SU(3)_2$ CS theory that contains the Fibonacci
anyon, (D) A fourth non-Abelian phase described by a 
$SU(3)_2 \times Z_2$ CS gauge theory. 
$m_1$, $m_2$ are phenomenological parameters
in the effective theory and drive the two types of Higgs transitions. They depend
on interlayer tunneling and inter/intra layer interactions in a way which requires further study. 
(C) and (D) both contain non-Abelian quasiparticles whose braiding statistics can realize universal TQC.
The minimal quasiparticle charge $e^*$ distinguishes (B) from the others. The others must be distinguished
in principle through detecting phase transitions in the neutral sector, or through tunneling/interferometry
measurements. 
}
\end{figure}

Several years ago \cite{barkeshli2010prl,barkeshli2011orb}, it was argued that the $(330)$ state,
in the presence of interlayer tunneling, could continuously transition to a different 
non-Abelian FQH state, known as the $Z_4$ Read-Rezayi FQH state \cite{read1999}, 
whose non-abelian braiding statistics alone is \it not \rm powerful enough for universal 
TQC 
Combining the earlier results with those 
of the present paper leads to a rich global phase diagram at total filling fraction $\nu = 2/3$ in bilayer systems, which we 
explore (see Fig. \ref{pdFig}). 

{\bf Thin torus limit--} For a wide variety of FQH states,
it was found\cite{seidel2005,seidel2006,bergholtz2006,bergholtz2008,seidel2008} that the wave 
function in the thin torus limit ($L_x/L_y\ll 1$) is smoothly connected to the fully 
two-dimensional wave function ($L_x/L_y \sim 1$), where $L_x$ and $L_y$ are the lengths 
of the torus in the two directions. This thin torus limit, which we review below,
allows for a simple understanding of fractionalization in the FQH state in terms of 
one-dimensional fractionalization \cite{heeger88}. 

In the $L_x/L_y \to 0$ limit and at filling fraction $1/n$, the dominant contribution to the pseudo-potential
Hamiltonian for the Laughlin state is
\begin{eqnarray}\label{Eq:thin-1}
&&H_{n} =\sum_{i} \sum_{0<r<n}U_{r,0} \hat{n}_{i}\hat{n}_{i+r},\quad U_{r,0}=g_{r,0}e^{-2\pi^2r^2/L_x^2}.
\end{eqnarray} 
$i$ indexes the lowest Landau level orbitals, extended in the $x$ direction and localized in the $y$ direction,
$g_{r,0}=1,r^2$ when $n=2,3$ respectively. Since $H_{n}$ involves only commuting 
number operators ($\hat{n}_i$'s), it can be immediately diagonalized. 
At $1/3$ filling, the following charge-density-wave patterns of electrons in the occupation basis minimize $H_3$:
$\ket{g}_{1}=\ket{100100100\cdots}$, $\ket{g}_{2}=\ket{010010010\cdots}$, $\ket{g}_{3}=\ket{001001001\cdots}$. 
In the two-dimensional limit, these three ground states evolve into the three topologically degenerate ground states 
of the $1/3$ Laughlin state on a torus \cite{bernevig2008,wen2008b,barkeshli2009}. 

The fractional quasiparticles can be understood as domain walls between 
these different patterns. For example, there is an excitation with $q=e/3$ charge at the domain wall between the
$[100]$ and $[010]$ patterns i.e. $[100][010]\equiv [\cdots1001001{\bf 00}|{\bf 0}10010\cdots]$, 
because there are three consecutive zeros, which leads to a deficit of charge $e/3$, according 
to the Su-Schrieffer counting \cite{heeger88}. The same is true for $[010][001]$ and $[001][100]$ 
patterns. In general, the domain wall between the ground states $|g\rangle_i$ and  
$|g \rangle_{(i+k) \% n}$ corresponds to a quasiparticle with electric charge $q = ke/n$. 

Now let us consider a double layer system, consisting of two identical layers,
in the presence of interlayer tunneling. In the thin torus limit, 
\begin{eqnarray}\label{Eq:H2}
H_{tt} = \sum_{i \atop 0<r<n} \left( U_{r,0}^{\alpha,\beta} \hat{n}_{i\alpha}\hat{n}_{i+r\beta} - 
t^{\perp}  c^{ \dagger}_{i\alpha} \sigma^x_{\alpha\beta} c_{i\beta}+H.c. \right)
\end{eqnarray} 
where $\alpha,\beta = \uparrow(\downarrow)$ refers to the 
top (bottom) layer, $\sigma^x = \left(\begin{matrix} 0 & 1 \\ 1 & 0 \end{matrix} \right)$,
$U_{r,0}^{\uparrow, \uparrow} = U_{r,0}^{\downarrow,\downarrow}$ and $U_{r,0}^{\uparrow,\downarrow} = U_{r,0}^{\downarrow, \uparrow}$ 
parametrize the intra- and inter- layer interactions, respectively. $H_{tt}$ is invariant 
under the $Z_2$ layer exchange symmetry $c_{i\uparrow} \leftrightarrow c_{i\downarrow}$. In the two-dimensional 
system, as long as interlayer tunneling $t^\perp$ is much smaller than the bulk gap in each layer, no phase transition 
is expected. As interlayer tunneling is increased, at some point the bulk gap can close and reopen 
in a different topological phase. In order to understand the resulting phase in a tractable limit,
we will study the effect of interlayer tunneling in (\ref{Eq:H2}). 

For simplicity, let us first consider only vertical tunneling, $t^\perp$ and ignore
the interlayer interactions, $U_{r,0}^{\uparrow\downarrow} = U_{r0}^{\downarrow\uparrow} = 0$. 
When $t^\perp = 0$, $H_{tt}$ has 9 exactly degenerate ground states, with the degeneracy protected by the
independent translation symmetries in each layer, and the layer exchange symmetry. 
When $t^\perp \neq 0$, the independent translation symmetries 
in each layer reduce to a single combined translation symmetry. The 3 $Z_2$ layer symmetric 
states, which we can label as $|D_1\rangle \equiv \left[\begin{array}{c}
    10 0 \\ 
   10 0\\ 
  \end{array}\right]$, 
$\ket{D_2}=\left[\begin{array}{c}
     010 \\ 
     010\\ 
  \end{array}\right]$, $\ket{D_3}=\left[\begin{array}{c}
     001 \\ 
     001\\ 
 \end{array}\right]$,
then acquire an energy splitting relative to the remaining 6 $Z_2$ layer symmetry-breaking 
ground states. As $t^\perp$ is increased further, we find that the energy gap closes,
and the 1D system passes through an Ising phase transition \cite{SM}. 
On the other side of the transition, there are 3 exactly degenerate ground states that are fully symmetric 
under the $Z_2$ layer exchange symmetry. Deep in this $Z_2$ symmetric phase, 
we can represent these states by a product over the state in each 3-site unit cell:
$|O_1\rangle \equiv \prod_{a =1}^{N_{uc}}
|\psi_1\rangle_a$, where $N_{uc}$ is the number of unit cells. 
Since $t^\perp_r \propto \delta_{r0}$, we have
\begin{align}
|\psi_1\rangle = \alpha_1  \left|\begin{array}{c}
     10 0 \\ 
     0 1 0\\ 
  \end{array}\right\rangle
 + \alpha_2  \left|\begin{array}{c}
     01 0 \\ 
     10 0\\ 
  \end{array}\right\rangle + \alpha_3  \left|\begin{array}{c}
     11 0 \\ 
     00 0\\ 
  \end{array}\right\rangle
+ \alpha_4  \left|\begin{array}{c}
     00 0 \\ 
     11 0\\ 
  \end{array}\right\rangle,
\end{align} 
where the other states are related by translations: 
$T_y |O_i \rangle = |O_{i+1} \rangle$. Here, $\alpha_j$, $j = 1,..,4$ 
are variational parameters, chosen to minimize the ground state energy. 

Therefore, for large enough interlayer tunneling, the 9 states that we started with split
into 3 degenerate $Z_2$ symmetric states $\{|O_i \rangle\}$, with energy $E_S$, 
3 degenerate states $\{|D_i\rangle\}$ with energy $E_D$, and 3 remaining degenerate 
$Z_2$ anti-symmetric states, with energy $E_A$. Now, we can consider two
distinct possibilities as we take the two-dimensional limit: either the 
6 states $\{|O_i \rangle, D_i\rangle\}$ continuously evolve into 
6 topologically degenerate ground states with a gap to other excited states,
or only 3 of the states (e.g. $\{ |O_i\rangle\}$), evolve into 3 topologically degenerate
ground states. Based on previous studies of the thin torus limit of FQH states \cite{seidel2005,seidel2006,bergholtz2006,bergholtz2008},
we expect that the former case will likely occur when $E_S \approx E_D \ll E_A$, 
while the latter case will occur in the regime $E_S \ll E_D, E_A$. Depending on 
parameters, $H_{tt}$ can access either regime; for example, $U^{\uparrow,\downarrow}_{0,0}<U^{\uparrow,\downarrow}_{1,0}$ or longer range tunneling can favor the former case over the latter. 

In what follows, we focus on the possibility where all six states, $\{|O_i\rangle$, $|D_i\rangle\}$, 
evolve into six topologically degenerate ground states in the 2D limit. 
The feasibility of this depends on microscopic details of the 2D system. This
appears to be a reasonable assumption because the results are in remarkable agreement 
with the effective field theory considerations presented below, and the earlier approach in \cite{mong2013,vaezi2013b}.
Additionally, the same assumption, when applied to the case of the (331) Halperin state, 
or the bosonic (220) state, yields results which agree with previous work
\cite{SM} \cite{read2000,wen2000,fradkin1999,seidel2008,papic2010,peterson2010}. 

It is natural to relabel the 6 ground states as follows:
$[200]$, $[020]$, $[002]$ denote $|D_i\rangle$, for $i = 1,2,3$, respectively,
and $[110]$, $[011]$, $[101]$, denote $|O_i \rangle$, for $i =1,2,3$. 
Below, our goal is to identify the type of topological order associated with this phase. 

First, observe that the total center of mass degeneracy (associated with translations $T_y$), only accounts for
a degeneracy of three. Therefore, the existence of $6$ states immediately signals the existence of a non-Abelian
FQH state. Recall that the quasiparticles can be understood as domain walls between the different ground state patterns. 
If we start with the state $[200]$ and consider a domain wall with the state $[110]$, then from the Su-Schrieffer
counting argument we see that there is a charge $e/3$ quasihole. This can be understood as the original 
Laughlin $e/3$ quasihole, but inserted in either the top layer or the bottom layer, with equal weight. 
If instead we start with the state $[110]$ and consider a domain wall with either $[020]$ or $[101]$, 
we see that there is again a charge $e/3$ quasihole. In general, we can ask which pairs of ground states, 
labelled $i$ and $j$, give rise to a charge $e/3$ quasihole at their domain wall. This defines an 
adjacency matrix for the charge $e/3$ quasihole,
\begin{eqnarray}\label{Eq:Adj-1}
 &&A=\para{\begin{array}{cccccc}
 0&0  &0 &1&0&0 \\
 0&0  &0 &0&0&1 \\
 0&0  &0 &0&1&0 \\
 0&1  &0 &0&1&0 \\
 1&0  &0 &0&0&1 \\
 0&0  &1 &1&0&0 \\
  \end{array}}
 \end{eqnarray}
where the rows/columns correspond to $[200],[020],[002],[110],[101],[011]$, respectively. 
More generally, let us consider $n_{qh}$ quasi-holes with $q=e/3$ at positions
$j_1,j_2,\cdots, j_{n_{qh}}$ \cite{ardonne2008}. To do so, we start with a fixed ground-state pattern, 
say $[200]$. At site $j_1$, there is a domain wall with $[110]$, at site $j_2$ 
there can be either $[020]$ or  $[101]$ patterns, and so on. We see that the number 
of possibilities grows exponentially with $n_{qh}$. It is straightforward 
to verify that ${\bf tr}\para{A^{n_{qh}}}$ gives the total number of different possibilities on the torus. 
Therefore, the degeneracy of the ground-state in the presence of $n_{qh}$ quasihole insertions 
grows as $\lambda_1^{n_{qh}}$ where $\lambda_{1}$ is the dominant eigenvalue of the adjacency 
matrix $A$. Consequently, the quantum dimension of the quasihole operator with minimum 
electric charge is $\lambda_1$. Using the above adjacency matrix, the quantum dimension of 
the charge $e/3$ quasihole is the golden ratio: $d_{qh}=F\equiv
\frac{1+\sqrt{5}}{2}$. 

Since there are six degenerate ground states on the torus, there are correspondingly 
six topologically distinct types of quasiparticles. These include the
$e/3$ quasiparticle described above and it's charge $-e/3$ particle-hole conjugate. 
There are also charge $2e/3$ and $4e/3$ quasiparticles, which involve inserting charge 
$e/3$ or $2e/3$ Laughlin quasiparticles into both layers simultaneously. We will label
them as $V_n$, with charge $q = 2n e/3$. These are inherited directly from 
the $(330)$ state, with their topological properties unchanged.
Finally, there is a neutral quasiparticle, which we label $\tau$. $\tau$
can be understood as inserting an $e/3$ quasiparticle
in the top layer and a $-e/3$ quasiparticle in the bottom layer, superposed 
with reverse process, $-e/3$ and $e/3$ in the top  and bottom layers, respectively. 
By studying the adjacency matrix, we find that the quantum dimension of 
$\tau$ is also $d_\tau = F$. 

The adjacency matrices $A_i$, for $i = 1,.., 6$, encode the fusion rules of the quasiparticles:
$i \times j = \sum_k (A_i)_{jk} k$, which dictates the number of ways quasiparticles $i$ and $j$ can
fuse into $k$ \cite{ardonne2008}. We find that the quasiparticles $V_a$ are simple Abelian quasiparticles, with
quantum dimension $1$: $V_a \times V_b = V_{a+b \% 3}$. Furthermore, 
$\tau \times \tau = 1 + \tau$; this is the fusion rule of the famous ``Fibonacci'' quasiparticle, whose
braiding statistics allows for universal topological quantum computation \cite{nayak2008}. 
The remaining two quasiparticles are identified with $V_a \tau$, for $a = 1, 2$. 

In addition to the fusion rules, we can obtain a information about the topological
spins. Since the theory has a subset of quasiparticles, 
$\{1, \tau\}$, with a closed fusion subalgebra $\tau \times \tau = 1 + \tau$, 
mathematical consistency \cite{ostrik2002} requires that the topological spin of $\tau$ be $\theta_\tau = \pm 2/5$.  
Furthermore, the quasiparticles $V_a$ are just the simple Abelian quasiparticles that were present
in the $(330)$ state. Since the phase transition occurs entirely within the neutral sector,
the topological spins of these charged quasiparticles should remain unchanged and
are given by their value in the $(330)$ state. Therefore, $\theta_{V_a} = a^2/3$. 
These results are summarized in Table \ref{Tab1}. 

\begin{table}
  \centering
  \begin{tabular}{@{} ccccc @{}}
    \toprule
     & Label & Charge (mod $e$) & Topological Spin & Quantum Dim.\\ 
   \hline 
    1& $V_0$ & $0$ & 0 &1\\ 
   2 &$V_1$ & $2e/3$& $1/3$ &1\\ 
   3 &$V_2$ & $e/3$& $1/3$&1\\ 
    4&$\tau$ & $0$& $\pm 2/5$ &F\\ 
    5&$V_1 \tau$ & $2e/3$ & $1/3 \pm 2/5$ &F\\ 
    6&$V_2\tau$ & $e/3$ & $1/3 \pm 2/5$ &F\\ 
    \hline 
  \end{tabular}
  \caption{The anyon content of the non-Abelian state obtained from the
$(330)$ state with strong interlayer tunneling. 
Plus (minus) sign denotes the two possibilities consistent with results obtained from the thin torus limit, and
correspond to the chirality of the non-Abelian sector, where the full edge theory has
central charge $c = 2 \pm 4/5$. The CS-Higgs theory fixes the $c = 14/5$ case. 
F is the golden ratio, $(1+\sqrt{5})/2$.}
  \label{Tab1}
\end{table}

Generalizing the above arguments to the $(nn0)$ states
gives $n(n+1)/2$ quasiparticles, whose fusion rules coincide
with the representation algebra of the quantum group $SU(n)_2$ \cite{SM}. Remarkably, the thin torus patterns 
$[200]$, $[020]$, $[002]$, $[110]$, $[011]$, $[101]$, and the connection to $SU(3)_2$,
have appeared previously in a completely different context \cite{ardonne2009}, in terms 
of the gapless, single-layer bosonic Gaffnian wave function. See also \cite{ardonne1999,hermanss2010} 
for other distinct realizations of $SU(3)_2$ fusion rules. 

The above results can be understood from the perspective of the edge
conformal field theory. Consider two free chiral bosons, $\varphi_1$ and $\varphi_2$, such that $e^{i n \varphi_1}$, 
$e^{i n \varphi_2}$ are considered to be local electron operators. In the $(nn0)$ state, 
$V_{a,b} \equiv e^{i a \varphi_1 + b \varphi_2}$, for $a,b = 0, ..., n-1$ correspond to the $n^2$ 
non-trivial quasiparticle operators. If we consider the $n(n+1)/2$ symmetrized operators
$\Phi_{a,b} = V_{a,b} + V_{b,a}$, and continue to treat the operators $e^{i n \varphi_1}$, 
$e^{i n \varphi_2}$ as trivial, local operators, then we find the remarkable result that
$\Phi_{a,b}$ satisfy the fusion rules of $SU(n)_2$: 
$\Phi_{a,b} \times \Phi_{a',b'} = \Phi_{a+a',b+b'} + \Phi_{a+b',b + a'}$. 
Recovering the topological spin from this procedure is more involved, as the stress-energy tensor in the CFT
also changes through this transition. 

{\bf Effective field theory--} Here, we show that there is a possible continuous phase 
transition between this non-Abelian FQH state and the $(330)$ state. We show that from 
the point of view of the effective field theory of the $(330)$ state, the appearance 
of the state we have found is quite natural in the presence of interlayer tunneling. 

One way of understanding the effective field theory of the $(330)$ state is through a 
parton construction \cite{wen1999}, where we write the electron operator as 
$c_\sigma = f_{1\sigma}f_{2\sigma}f_{3\sigma}$, 
where $\sigma = \uparrow, \downarrow$ is the layer index, and $f_{i\sigma}$ are charge $e/3$ fermionic `partons.' 
This rewriting of the electron operator introduces an $SU(3) \times SU(3)$ gauge symmetry, associated 
with the transformations $f_\sigma \rightarrow W_\sigma f_\sigma$, for $W_\sigma \in SU(3)$, which keep all 
physical operators invariant. The theory in terms of electron operators can therefore be replaced by a theory in 
terms of the partons $f_{a\sigma}$, coupled to an $SU(3)$ gauge field, $A_\sigma$. In the presence of a magnetic
field $B$, the partons feel an effective magnetic field $B_{eff} = B/3$. When the electrons are at filling $1/3$, 
the partons are then poised to form a $\nu = 1$ integer quantum Hall state at mean-field level. 
Integrating out the partons then gives an $SU(3)_1 \times SU(3)_1$ CS gauge theory:
$\mathcal{L}=\frac{\epsilon^{\mu\nu\lambda}}{4\pi} \sum_\sigma {\rm tr} \para{ A^\sigma_\mu \partial_\nu A^\sigma_\lambda + \frac{2}{3} A^\sigma_\mu A^\sigma_\nu A^\sigma_\lambda} + j_\sigma \cdot A^\sigma$.
$j_\sigma$ is the current of quasiparticles, which, after integrating out the partons, appear in this 
theory as classical `test' charges. They correspond to the fermionic particles/holes in the parton Landau levels,
and acquire fractional statistics after being dressed by the CS gauge field. 

Next, let us consider the effect of interlayer tunneling, 
$\delta \mathcal{H}_{t} = -t^\perp c_\uparrow^\dagger c_\downarrow + h.c. 
= -t^\perp (f_{1\uparrow}f_{2\uparrow}f_{3\uparrow})^\dagger f_{1\downarrow}f_{2\downarrow}f_{3\downarrow} + H.c.$, 
on the mean-field state of the partons. For $t^\perp$ large enough, this induces a non-zero expectation value:
$\langle f_\uparrow^\dagger f_\downarrow\rangle \neq 0$, which breaks the gauge symmetry $SU(3) \times SU(3) \rightarrow SU(3)$,
leaving a single gauge field $A \equiv A^\uparrow = A^\downarrow$ at long wavelengths. Now, integrating out the
partons leads to a $SU(3)_2$ CS gauge field: 
$\mathcal{L}_{CS,\sigma}=\frac{2}{4\pi} \epsilon^{\mu\nu\lambda} {\rm tr} \para{ A_\mu \partial_\nu A_\lambda + \frac{2}{3} A_\mu A_\nu A_\lambda}$.
At the critical point, only the fluctuations of the electrically neutral operator $f_\uparrow^\dagger f_\downarrow$ will be massless.
Consequently, charged fluctuations remain gapped across the transition.

The edge CFT of the parton mean field states is described by a $U(6)_1$ chiral Wess-Zumino-Witten 
CFT. Implementing the projection onto the physical degrees of freedom yields a $U(6)_1/SU(3)_2$ coset theory, with 
central charge $c = 14/5$. We can systematically obtain the topological properties of the 
quasiparticles in this theory \cite{SM}. Remarkably, the result coincides with the $SU(3)_2$ fusion rules obtained 
from the thin torus limit above, and the topological spins match those of Table \ref{Tab1} exactly, 
with the choice $\theta_\tau = 2/5$. We conclude that there exists a continous phase transition 
between these two phases, associated with the Chern-Simons-Higgs transition 
$SU(3)_1 \times SU(3)_1 \rightarrow SU(3)_2$. The generalization to $(nn0)$ states gives
$SU(n)_1 \times SU(n)_1 \rightarrow SU(n)_2$ CS-Higgs transitions, all of which 
match results obtained from symmetrizing the thin torus patterns. The case
$n = 2$ is related to \cite{zhang2014}; it is closely related to, but distinct from,
the theory of \cite{fradkin1998,fradkin1999}, since the edge theory of the non-Abelian state in this case
is $U(4)_1/SU(2)_2 \neq SU(2)_2$. 

The parton construction suggests wave functions that capture the universal
features of this state. In a continuum system, a natural ansatz is 
$\mathcal{P}_{LLL}(\Phi_{\nu  =2})^3$, where $\Phi_{\nu =2}$ is a wave function where the two lowest symmetric 
Landau levels are filled, and $\mathcal{P}_{LLL}$ is the projection onto the lowest Landau level. 
On a lattice, one can consider $\Phi_{C = 2}(\{r_i\})^3$ \cite{zhang2013}, 
where $\Phi_{C = 2}(\{r_i\})$ is a wave function for a band insulator
with Chern number 2. 

{\bf Global phase diagram}--\rm The above field theoretic understanding
helps us understand the relation of this non-Abelian state to the $Z_4$ RR state, which 
can also continuously transition to the $(330)$ state \cite{barkeshli2010prl,rezayi2010}. As was shown in
\cite{barkeshli2010,barkeshli2011orb}, the $Z_4$ Read-Rezayi state can be understood 
in terms of $[SU(3)_1 \times SU(3)_1] \rtimes Z_2$ CS gauge theory.
Here, the meaning of the $\rtimes Z_2$ is that the symmetry of interchanging the two $SU(3)$ gauge
fields is itself promoted to a local gauge symmetry. The transition from the $Z_4$ RR state
to the $(330)$ state can be understood as a $Z_2$ gauge symmetry breaking transition:
$[SU(3)_1 \times SU(3)_1] \rtimes Z_2 \rightarrow SU(3)_1 \times SU(3)_1$. Combining this
with the result above, we see that there are four closely related phases that are separated
by continuous phase transitions (see Fig. \ref{pdFig}). 

Returning to the thin torus Hamiltonian (\ref{Eq:H2}), in the limit
where $t^\perp$ is the largest energy scale the electrons only occupy the symmetric
orbitals on each site, with two electrons per unit cell. In this limit the ground state 
will be 3-fold degenerate. Since this degeneracy can be accounted 
for by center of mass translations $T_y$, the resulting state is Abelian and corresponds to 
the particle-hole conjugate of the $1/3$ Laughlin state. A similar result is obtained 
in the context of the $(331)$ state \cite{papic2010}.

\bf{Conclusion}\rm-- At the transition between the $(330)$ state and the non-Abelian states,
the charge gap remains open while the neutral gap closes. Past experiments \cite{suen1994,manoharan1996,lay1997},
which have probed the $\nu = 2/3$ phase diagram in bilayers through
resistivity measurements, were directly sensitive only to the charge gap and thus
have not yet definitively ruled out these exotic non-Abelian states and transitions. 

In the Supplemental Materials, we discuss a different ``coupled wire''
approach and show the remarkable agreement with the results presented
above, we provide additional details and generalizations of our analyses, and 
we discuss the duality between $(nnl)$ bilayer state with interlayer pairing and $(n,n,-l)$ state with interlayer tunneling.

\bf{Acknowledgments}\rm-- This work was supported by the Bethe postdoctoral fellowship (AV) and Microsoft Station Q (MB).
We thank K. Lee, C. Nayak, P. Bonderson, J. Shabani, A. Stern, E. Berg, A. Ludwig, Z. Wang, A. Seidel, J.C. Davis, E. Mueller, J. Alicea,
D. Clarke, R. Mong, X.-G. Wen for discussions, and E.-A. Kim for helpful comments on the manuscript.

%

\newpage
\appendix

\begin{widetext}

\section*{Supplemental materials}

\section{1. Coupled quantum wires approach}

Here, we study the $(330)$ Halperin bilayer state in the presence of strong interlayer tunneling from a 
different perspective, adapting the coupled-wire approach of \cite{mong2013,vaezi2013,teo2011} to the case at hand. 
Remarkably, this approach yields results that match those presented from completely different perspectives
in the main text.

Let us consider two adjacent (330) bilayer FQH bars. At their interface, we have two left-moving chiral modes
from the upper bar, and two right-moving anti-chiral modes on the lower bar (see Fig. \ref{fig:Fig2}). 
The electron destruction operator on each of these modes is:
\begin{eqnarray}
&&\Psi_{1,R} \propto e^{i\sqrt{3}\phi_{1,R}} \equiv e^{ i{\sqrt{\frac{3}{2}}}\para{\phi_{c,R}+\phi_{s,R}}} \cr
&&\Psi_{2,R} \propto e^{i\sqrt{3}\phi_{2,R}} \equiv e^{ i{\sqrt{\frac{3}{2}}}\para{\phi_{c,R}-\phi_{s,R}}},
\end{eqnarray}
and similarly for the left-moving modes. $\phi_{I,R}$, for $I = 1,2$ is the right-moving boson on the 
$I$th layer, while $\phi_{I,L}$ is the left-moving boson from the $I$th layer. $\phi_{I,R}$ and $\phi_{I,L}$ 
are compactified on a circle of radius $R = \sqrt{3}$: $\phi_{I,R} \sim \phi_{I,R} + 2\pi \sqrt{3}$, and similarly
for $\phi_{I,L}$. It is convenient to define the linear combinations:
\begin{align}
\phi_{cR} &= \frac{1}{\sqrt{2}} (\phi_{1,R} + \phi_{2,R}),
\nonumber \\
\phi_{sR} &= \frac{1}{\sqrt{2}} (\phi_{1,R} - \phi_{2,R}),
\end{align}
and similarly for the left movers. These describe the charged and neutral modes, respectively. 
The Hamiltonian that describes these four gapless modes in the absence of any perturbation is:
\begin{eqnarray}
\mathcal{H}_{0} =\sum_{\tau=c,s} \frac{1}{4\pi}\int dx \para{\para{\partial_x \varphi_{\tau}}^2+\para{\partial_x \theta_{\tau}}^2}.
\end{eqnarray}
where $\varphi_{c/s}=\frac{\phi_{c/s,R}+\phi_{c/s,L}}{\sqrt{2}}$, $\theta_{c/s}=\frac{\phi_{c/s,R}-\phi_{c/s,L}}{\sqrt{2}}$ 
are conjugate bosonic variables. These new bosons are all compactified on a circle with radius
$R=\sqrt{3}$. 

\begin{figure}
\includegraphics[width=.5\textwidth]{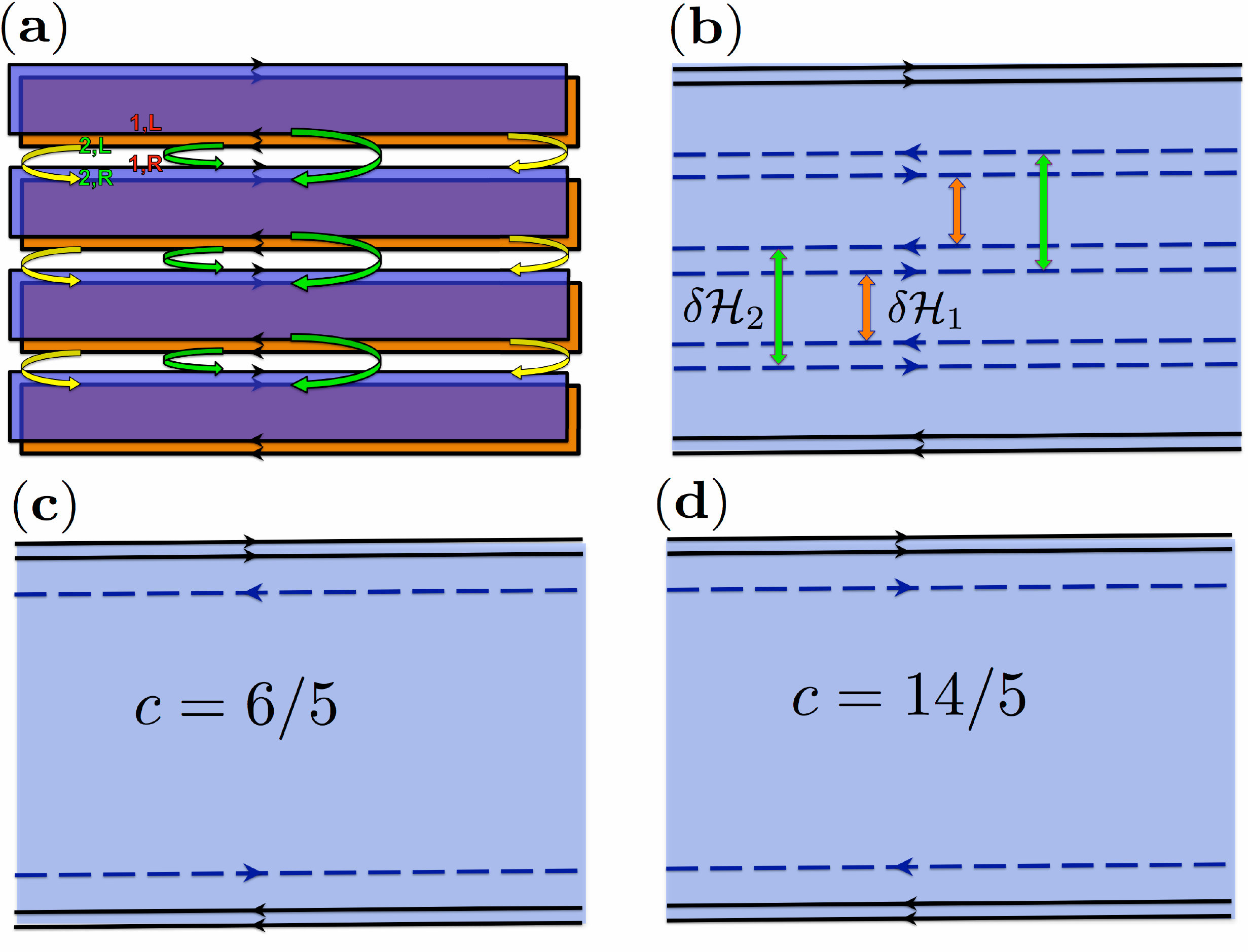}
\caption{Coupled wire construction in $(330)$ state. (a) The (yellow) green arrows 
represent the (intra-wire) inter-wire coupling. Note that the outermost free boson modes 
do not couple to other modes, and remain gapless. (b) By tuning the 
strengths of the two different types of backscattering terms, we obtain counterpropagating 
gapless $Z_3$ parafermion modes, shown in dotted lines. We have drawn the right and 
left moving parafermion modes to be spatially separated, although their spatial profile
may be more complicated. The parafermion chains can in principle couple in two different 
ways, shown pictorially by the blue and orange arrows. These couplings may require strong electron tunneling,
in addition to allowed quasiparticle tunnelings, and can gap the counterpropagating modes. (c) When the strength of orange 
type $(\delta \mathcal{H}_1)$ inter-wire coupling dominates, the topmost anti-chiral parafermion 
mode remains gapless. This case corresponds to $c=6/5$ total central charge of the chiral edge 
CFT. (d) When the strength of green type inter-wire coupling $(\delta \mathcal{H}_2)$ dominates, 
the topmost chiral parafermion remains gapless. This case corresponds to $c=14/5$ total 
central charge of the chiral edge CFT.}
\end{figure}\label{fig:Fig2}

Next, we consider the following perturbations, corresponding to intra-layer or inter-layer backscattering between counter-propagating
edge states:
\begin{eqnarray}
\delta \mathcal{H}_{\parallel} = && -t_{\parallel}\para{\Psi_{1,R}^\dag \Psi_{1,L}+\Psi_{2,R}^\dag \Psi_{2,L} }+H.c.\cr
\delta \mathcal{H}_{\perp} =&&-t_{\perp}\para{ \Psi_{1,R}^\dag \Psi_{2,L}+\Psi_{2,R}^\dag \Psi_{1,L}}+H.c.
\end{eqnarray}
In terms of the bosonic variables, we have:
\begin{eqnarray}
\delta \mathcal{H} = -4 \cos\para{\sqrt{3}\theta_{c}}\para{t_{\perp}\cos\para{\sqrt{3} \varphi_{s}}+t_{\parallel}\cos\para{\sqrt{3}\theta_{s}}}.
\end{eqnarray}
Since $\theta_{c}\para{x}$ commutes with every term in the Hamiltonian, we can condense it and 
replace $\cos\para{\sqrt{3}\theta_{c}}$ with its expectation value. Thus we obtain the following effective Hamiltonian:
\begin{eqnarray}
\mathcal{H}_{\rm eff} =\frac{1}{4\pi} && \int dx \left[\para{\partial_x \varphi_{s}}^2+\para{\partial_x \theta_{s}}^2\right]\cr
-u&& \int dx ~\left[t_{\perp} \cos\para{\sqrt{3} \varphi_{s}}+t_{\parallel} \cos\para{\sqrt{3}\theta_{s}} \right].
\end{eqnarray}
where $u=4\braket{\cos\para{\sqrt{3}\theta_c}}$. The above effective Hamiltonian is the 
well studied $\beta^2=6\pi$ self-dual sine-Gordon model, because the scaling dimension 
of the perturbations is $3/2$ \cite{lecheminant2002}. This sine-Gordon model describes the 
low energy physic of a $Z_3$ quantum clock model and $Z_3$ parafermion chain. Moreover, 
its self-dual point ($t_{\perp}=t_{\parallel}$) is described by a $Z_3$ parafermion CFT whose 
chiral (anti-chiral) sector has $c=4/5$ ($c=-4/5$) central charge. 

Next, following \cite{mong2013,teo2011,vaezi2013} we consider a 2D array of the 1D wires, 
with proper inter-wire couplings. Each of the 1D wires has counterpropagating $Z_3$ parafermion
modes. Depending on the types of quasiparticle and electron tunnelings between the different
parafermion wires, two scenarios are possible in principle. In one, 
it is possible to gap out the left-moving mode on each wire with the right-moving
mode on the wire below it, leaving a right-moving parafermion mode on the top-most wire, and a left-moving
parafermion mode on the bottom-most wire, while the interior of the system is fully gapped. Alternatively,
we may consider the case where the right-moving mode on each wire is gapped with the left-moving mode on the
wire below it. This scenario leaves a left-moving $Z_3$ parafermion mode on the top-most wire,
and a right-moving mode on the bottom-most wire. 

The outer edge of the sample, which is the edge between the $(330)$ state and vacuum, 
contains two right-moving bosonic modes with central charge $c = 2$. The two scenarios 
outlined above give rise to an additional $Z_3$ parafermion mode, with central charge
$c = \pm 4/5$. Therefore the total chiral central charge of the state is $c = 2 \pm 4/5$. 
We see that this agrees with results obtained from the thin torus limit. The case $c = 2 + 4/5$
agrees also with the results from the CS-Higgs theory presented in the main text, and with
more detailed results obtained in \cite{mong2013}. 

\section{2. Thin Torus Results}

\subsection{a. Brief introduction to thin torus limit}

The single-particle wavefunction of the lowest Landau levels on the torus in the Landau gauge , ${\bf A}=\para{-By,0}$ is
\begin{eqnarray}
&&\Psi_{m}\para{x,y}=\sum_{n} e^{i\para{nL_y+\frac{2\pi}{L_x}m}x}e^{-\para{y+nL_y+\frac{2\pi m}{L_x}}^2 /2},
\end{eqnarray}
for $\hbar=c=e=B=1$. In these units, the magnetic length and the number of states in the lowest Landau level (LLL) are   $l_{B}=\sqrt{\hbar c/eB}=1$, and $N_{f}=\frac{L_xL_y}{2\pi}$, respectively. The wave function $\Psi_{m}$ is extended along the $x$ direction with momentum $k_x=2\pi m/L_x$ (mod $ L_y$) and localized along the $y$ direction around $y_{0}\para{m}=-\frac{2\pi}{L_x}m$ (mod $L_y$). Hence, the average distance between the peaks of two successive wave functions is $2\pi/L_x$, while the average width of the wave function along $y$ direction is of order unity. Thus, when $L_x <<1$, the overlap between two adjacent wave functions, $\Psi_{m}$ and $\Psi_{m+1}$, is negligible. 

Now consider the fractional quantum Hall (FQH) problem at filling fraction $\nu=N_e/N_f=1/n$ on the torus, 
which has $n$ degenerate many body ground-states. As long as the topological properties are concerned, 
we can study the following model Hamiltonian
\begin{eqnarray}\label{Eq:H1}
H_{n}=\sum _{i} \sum_{r>s} U_{r,s} c_{i+s}^\dag c_{i+r}^\dag c_{i+r+s} c_{i},
\end{eqnarray} 
where $c_{m}^\dag$ denotes the electron creation operator with $\Psi_{m}\para{x,y}$ wave function and 
\begin{eqnarray}\label{Eq:int-1}
U_{r,s}=g_{r,s}e^{-2\pi^2\para{r^2+s^2}/L_x^2}.
\end{eqnarray} 
where $g_{r,s}=1$ for $n=2$ and $g_{r,s}=\para{r^2-s^2}$ for $n=3$. The above Hamiltonian has $n$ degenerate ground-states described by the Laughlin wave-functions on the torus. 

\section{b. Ising phase transition in thin torus limit for (330) state with tunneling}

Consider the thin torus Hamiltonian for the $(330)$ state with interlayer tunneling, displayed in 
eq. (\ref{Eq:H2}) of the main text, for the case where the interlayer interactions 
$U^{\uparrow\downarrow}_{r,0}=U^{\downarrow\uparrow}_{r,0} = 0$:
Next consider the following thin torus Hamiltonian:
\begin{eqnarray}
&&H= - \sum_{i} t^{\perp} c_{i,\up}^\dag c_{i,\dn}+h.c.+\sum_{i,\sigma=\up,\dn}\para{U_{1,0}n^{\sigma}_{i}n^{\sigma}_{i-1}+U_{2,0}n^{\sigma}_{i}n^{\sigma}_{i-2}}.
\end{eqnarray}
We claimed that as $t^{\perp}$ is increased from 0, eventually the thin torus system undergoes an Ising
phase transition. Here we provide the details of this analysis. 

To see this, first recall that for small $t^\perp$, the diagonal states $|D_i\rangle$, for $i = 1,2,3$, 
defined in the main text, acquire an energy splitting relative to the other degenerate 6 states. The degeneracy
of 6 is protected by the spontaneously broken translation symmetry, along with the spontaneously broken
$Z_2$ interlayer exchange symmetry. 

To make progress, let us pick one of these six degenerate ground states, when $t^\perp = 0$, say
\begin{align}
\prod_{i=1}^{N_{uc}} \left| \begin{array}{ccc}
    1 & 0 & 0  \\ 
    0 & 1 & 0 \\ \end{array} \right\rangle_i,
\end{align}
where $N_{uc}$ is the number of 3-site unit cells. 
Since the interlayer tunneling is purely vertical, we can analyze the consequence of $t^\perp$ by considering 
the following states within each unit cell:
\begin{eqnarray}
&&\left|  \begin{array}{ccc}
    1 & 0 & 0  \\ 
    0 & 1 & 0 \\ \end{array} \right\rangle_i ,\quad
\left| \begin{array}{ccc}
    0 & 1& 0  \\ 
    1& 0 & 0 \\  \end{array} \right\rangle_i ,\quad 
\left| \begin{array}{ccc}
    1 & 1 & 0  \\ 
    0 & 0 & 0 \\  \end{array} \right\rangle_i,\quad
\left|\begin{array}{ccc}
    0 & 0 & 0  \\ 
    1 & 1 & 0 \\  \end{array}\right\rangle_i.
\end{eqnarray}
Here, $i$ labels the $i$th 3-site unit cell. 

We now assume that $U_{1,0}$ is the largest energy scale in the Hamiltonian. Therefore, the following 
two local states $\left| \begin{array}{ccc}
    1 & 1 & 0  \\ 
    0 & 0 & 0 \\  \end{array} \right\rangle_i$ and $\left|\begin{array}{ccc}
    0 & 0 & 0  \\ 
    1 & 1 & 0 \\  \end{array}\right\rangle_i$are costly, with energies of order $U_{1,0}$.
Thus we can integrate them out in the low energy limit. Therefore, we only consider the 
following two states and represent them by a single spin 1/2 per unit cell:
\begin{eqnarray}
&&\left|\begin{array}{ccc}
    1 & 0 & 0  \\ 
    0 & 1 & 0 \\ \end{array} \right\rangle_i \equiv \ket{\up}_i,\quad 
    \left|\begin{array}{ccc}
    0 & 1& 0  \\ 
    1& 0 & 0 \\  \end{array} \right\rangle_i\equiv \ket{\dn}_i
\end{eqnarray}
Using second order perturbation theory, the effective spin Hamiltonian in the limit $t^{\perp} \ll U_{1,0}$ is:
\begin{eqnarray}\label{eq:Ising-1}
H_{\rm eff}=-J\sum_{i}S_{i}^{x}-\frac{U_{2,0}}{2}\sum_{i} \para{S^z_{i}S^z_{i+1}+1}, \quad J=4\para{t^{\perp}}^2/U_{1,0}.
\end{eqnarray}

The above Ising chain has two gapped phases: a ferromagnetic phase for $J<U_{2,0}/2$, and a paramagnetic phase 
for $J>U_{2,0}/2$. Physically the $Z_2$ symmetry is that of the exchanging the two layers. 
The ferromagnetic phase is doubly degenerate; deep in the phase ($J/U_{2,0} \rightarrow 0$) the 
degenerates eigenstates are of the form $\prod_{i}\ket{\up}_i$, and $\prod_{i}\ket{\dn}_i$. 
Note that none of these ground-states are $Z_2$ symmetric, but rather they transform into each other under the $Z_2$ symmetry. 
The paramagnetic phase has a unique $Z_2$ symmetric ground-state. Deep in this phase ($J/U_{2,0} \rightarrow \infty$), the
spins are aligned along the $x$ direction, and the ground state is $\prod_i \frac{\ket{\up}_i+\ket{\dn}_i}{\sqrt{2}}$. 

The effective spin Hamiltonian in Eq. \eqref{eq:Ising-1} can be exactly solved by the Jordan-Wigner transformation. 
In this method, we can represent $S_{i}^{x}$ and $S_{i}^zS_{i+1}^z$ using two Majorana fermions per site as follows: 
\begin{eqnarray}
&& S_{i}^{x} = i \gamma^{1}_{i}\gamma^{2}_{i}, \quad  S_{i}^{z}S_{i+1}^{z} = i \gamma^{2}_{i}\gamma^{1}_{i+1}.
\end{eqnarray}
We can combine the two flavors of the Majorana fermions to obtain a single complex fermion per site:
\begin{eqnarray}
\gamma^{1}_i=\para{f_{i}+f_{i}^\dag}, \quad \gamma^2_i=-i\para{f_{i}-f_{i}^\dag}  
\end{eqnarray}
After the above transformations, we obtain the Kitaev 1D chain, which is equivalent to a p-wave superconductor. 
We can easily diagonalize the block Hamiltonian associated with this superconductor and obtain the full spectrum. 
The fermion excitation gap is $\abs{2J-U_{2,0}}$, and therefore the Ising transition occurs when the gap closes: $J=U_{2,0}/2$.

\section{3. $SU(3)_2$ parton construction, and $U(6)_1/SU(3)_2$ edge theory}

In this section we will discuss the topological properties of the $SU(3)_2$ parton construction presented in the
main text, in order to establish the remarkable agreement with the results of the thin torus limit. 
It will be easiest to discuss this in terms of the edge theory, which is a $U(6)_1/SU(3)_2$ coset theory. 

To make progress, let us first review the properties of the pure $SU(3)_2$ WZW theory \cite{difrancesco}. 
The central charge is $c_{SU(3)_2} = \frac{16}{5}$. 
The primary fields are described by the integrable highest weight representations 
of the $SU(3)_2$ affine Lie algebra. There are 6 integrable highest weight 
representations; the $SU(3)$ weights for these representations can be labelled as
$(2,0)$, $(0,2)$, $(0,0)$, $(1,1)$, $(1,0)$, and $(1,1)$. The scaling dimensions 
are given by the following formula:
\begin{align}
h_{\lambda} = \frac{(\lambda,\lambda+ 2\rho)}{2(k+g)}, 
\end{align} 
where in our case $k  =2$ and $g = 3$, and $\rho = (1,1)$ is the Weyl vector. Denoting
\begin{align}
\omega_1 = (1,0), \;\; \omega_2 = (0,1),
\end{align}
these have the inner products:
\begin{align}
(\omega_1,\omega_1) = (\omega_2,\omega_2) = 2/3, \;\; (\omega_1,\omega_2) = 1/3. 
\end{align}
Using these inner products we can obtain all of the scaling dimensions:
\begin{align}
h_{(0,0)} &= 0
\nonumber \\
h_{(1,0)} &= \frac{(\omega_1, 3\omega_1 + 2\omega_2)}{10} = \frac{4}{15}
\nonumber \\
h_{(0,1)} &= \frac{(\omega_2, 3\omega_2 + 2\omega_1}{10} = \frac{4}{15}
\nonumber \\
h_{(1,1)} &= \frac{3(\omega_1 + \omega_2, \omega_1 + \omega_2)}{10} = 3/5
\nonumber \\
h_{(2,0)} &= \frac{ 4(\omega_1, 2\omega_1 + \omega_2)}{10} = 2/3
\nonumber \\
h_{(0,2)} &= \frac{4(\omega_2, \omega_1 + 2\omega_2)}{10} = 2/3
\end{align}

Now let us consider the theory of interest, $U(6)_1/SU(3)_2$, which is the edge theory for 
our $SU(3)_2$ CS theory coupled to fermionic partons. The central charge is
\begin{align}
c = c_{U(6)_1} - c_{SU(3)_2} = 6 - 16/5 = 14/5
\end{align}
The stress tensor is given by
\begin{align}
T_{U(6)_1/SU(3)_2} = T_{U(6)_1} - T_{SU(3)_2},
\end{align}
which implies that our six primary fields will actually obtain the following scaling dimensions:
\begin{align}
h_{(0,0)} = 0,
\nonumber \\
h_{(1,0)} = 1/2 - 4/15 = 7/30
\nonumber \\
h_{(0,1)} = 1/2 - 4/15 = 7/30
\nonumber \\
h_{(1,1)} = 1 - 3/5 = 2/5
\nonumber\\
h_{(2,0)} = 1 - 2/3 = 1/3
\nonumber \\
h_{(0,2)} = 1 - 2/3 = 1/3
\end{align}
In terms of the CS parton theory, $\Phi_{(1,0)}$ and $\Phi_{(0,1)}$ can be understood
as an odd number of holes in the parton bands, dressed by the $SU(3)_2$ CS gauge field.
In contrast, $\Phi_{(1,1)}$, $\Phi_{(2,0)}$, and $\Phi_{(0,2)}$ can be understood as
an even number of holes in the parton bands, dressed by the $SU(3)_2$ CS gauge field. 
Remarkably, we see that these scaling dimensions and the central charge $c = 4/5$ exactly 
match the results of Table \ref{Tab1} in the main text (note that the topological spins in 
Table \ref{Tab1} are defined only modulo $1/2$, because of the existence of local fermionic
electrons in the system). Furthermore, note that these operators all obey the $SU(3)_2$ fusion rules.
\bf{Therefore, the $SU(3)_2$ parton construction along with the $U(6)_1/SU(3)_2$ edge theory
exactly reproduce results obtained from the thin torus limit!}\rm. 

To conclude, we observe that the edge CFT can also be understood in terms of a
$SU(2)_3 \times U(1)_6$ CFT, where the electron operator is defined as $\Phi^3_0 e^{i \sqrt{3/2} \phi_c}$. 
Here, $\phi_c$ is a chiral boson representing the charge sector, and $\Phi^n_0$ is the spin-$n/2$ 
primary field from the $SU(2)_3$ theory. Given this electron operator, the quasiparticle operators
consist of those operators which are mutually local with respect to the electron operator. These, and their
scaling dimensions, are:
\begin{align}
\mathbb{I} , &\;\;\; h = 0
\nonumber \\
e^{i2/3 \sqrt{3/2} \phi_c}, &\;\;\; h = 1/3
\nonumber \\
e^{i 4/3 \sqrt{3/2} \phi_c}, &\;\;\; h = 4/3 
\nonumber \\
\Phi^1_0 e^{i1/3 \sqrt{3/2} \phi_c}, &\;\;\; h = 3/20 + 1/12 = 7/30
\nonumber \\
\Phi^2_0, &\;\;\; h = 2/5
\nonumber \\
\Phi^1_0 e^{i 5/3 \sqrt{3/2} \phi_c}, &\;\;\; h = 3/20 + 25/12 = 1 + 7/30
\end{align}
We see that modulo 1, these are exactly the same scaling dimensions as we found
from the $U(6)_1/SU(3)_2$ coset theory above! Actually, this should be expected, because of the identity:
\begin{align}
U(k n)_1 = SU(n)_k \times SU(k)_n \times U(1)_{kn},
\end{align} 
which, when applied to our case, gives
\begin{align}
U(6)_1/SU(3)_2 = SU(2)_3 \times U(1)_6. 
\end{align}

{Before closing this section, we would like to mention that the famous Gauss-Milgram relation cannot be applied to Table I of the main text (with positive sign) in order to find the central charge (mod 8). The reason is that Gauss-Milgram theorem applies only to bosonic topological phases, whereas in this paper we are considering a topological state where the microscopic constituents are fermions (electrons). More precisely, the Gauss-Milgram theorem applies to modular topological quantum field theories, while phases built out of fermions are not modular. As a simple example, consider the 1/3 Laughlin state (which is a simpler "fermionic" topological phase). The topological spins are $0, \pi/3, \pi/3$, and are well-defined only modulo $\pi$, because the electron has fermionic statistics and is topologically trivial. Applying the Gauss-Milgram sum  gives nonsense in general (the LHS isn't even a pure phase), unless we make the choices $0, 4\pi/3, 4\pi/3$, in which case we would incorrectly get a central charge of 2. The correct central charge for the 1/3 Laughlin state is 1. \\ 

Since the Gauss-Milgram sum does not apply in our theory, we use additional input from the other approaches in order to fix the central charge. The coupled wire construction (discussed in the Supplemental Material), predicts a central charge c = 2 $\pm 4/5$. The CS Higgs theory fixed the central charge c = 14/5. }\\

\section{4. Generalizations: $SU(n)_1 \times ...\times SU(n)_1 \rightarrow SU(n)_k$}

So far we have mainly focused on the special case of the $(330)$ states, which can transition to
the exotic $SU(3)_2$ non-abelian states as a function of interlayer tunneling, as these are most
relevant to experimentally accessible systems. However the above results 
generalize to a much wider class of examples. In general, we can consider $k$ layers
of $1/n$ Laughlin states in each layer, and consider interlayer tunneling among all of the layers. Repeating
the arguments from the thin torus limit in this more general case gives $SU(n)_k$ fusion rules. Additionally,
the effective field theory construction presented in the main text naturally generalizes to $SU(n)_k$ CS theory,
with the edge theory described by $U(n)_1/SU(n)_k$ coset CFT. 

The special case of $n = 2$ is therefore closely related, but not identical to, the results of 
previous studies \cite{fradkin1998,fradkin1999}. 
In particular, the case of two $1/2$ Laughlin states, \it i.e. \rm the $(220)$ states gives, 
$SU(2)_2$ fusion rules. 
Below, we will discuss the $(220)$, and then the general $(nnn0)$ case in some more detail. 

\subsection{a. $(220)$ results}

Let us consider the bosonic $(220)$ bilayer FQH state in the strong interlayer tunneling regime, and 
compute the quantum dimension of its quasihole excitation. It is known \cite{fradkin1999,read2000,wen2000} 
that there is a phase transition into the bosonic Moore-Read state \cite{moore1991}, whose quasihole 
excitation is related to Ising anyon and therefore has quantum dimension $d=\sqrt{2}$. Below we will recover this
result by adapting the thin torus argument presented in the main text. 

In the absence of tunneling, each layer has two degenerate ground-states:  
$\ket{g}_{1}=\ket{101010\cdots}:= \left[10\right]$, 
$\ket{g}_{2}=\ket{010101\cdots}:= \left[01\right]$. Therefore, the whole 
system has the following degenerate ground-states: $\ket{g}_{1,1}=\left[\begin{array}{c}
     10  \\ 
     10 \\ 
  \end{array}\right]$, $\ket{g}_{1,2}=\left[\begin{array}{c}
     10  \\ 
     01 \\ 
  \end{array}\right]$, $\ket{g}_{2,1}=\left[\begin{array}{c}
     01  \\ 
     10 \\ 
  \end{array}\right]$, and $\ket{g}_{2,2}=\left[\begin{array}{c}
     01  \\ 
     01 \\ 
  \end{array}\right]$. As explained in the main text, for strong enough  inter-layer tunneling, the
system can pass through a phase transition into a new phase, where 
we keep only the symmetrized ground states. In this case, this consists of 3 states:
$\ket{\alpha}_1=\ket{g}_{1,1}$, $\ket{\alpha}_2=\ket{g}_{2,2}$, and $\ket{\alpha}_3=\left[\para{\begin{array}{c}
     10  \\ 
     01 \\ 
  \end{array}}+\para{\begin{array}{c}
     01  \\ 
     10 \\ 
  \end{array}}\right]$. We can represent these three degenerate states by $\left[20\right]$, 
$\left[02\right]$, and $\left[11\right]$ by summing over the occupations of each layer. 

Since, the total filling fraction is one, the total center of mass momentum does not contribute to 
the ground-state degeneracy; the ground state degeneracy of $3$ therefore signals the 
non-Abelian nature of the resulting state. Let us compute the fusion rules of the charge $e/2$
quasiparticle by computing the adjacency matrix, as explained in the main text. 
Note that because we assume the bosons have charge $e$, charge $e/2$ and charge $-e/2$ are actually
topologically equivalent excitations because they differ by a local operator. 
We consider the ways of creating the charge $q = e/2$ quasihole by creating domain walls between
the different occupation number patterns. We see that a domain wall between 
$[20]$ and $[11]$ localizes a charge $e/2$ quasihole. Similarly, $[02]$ followed by $[11]$ also creates
the same type of excitation (modulo the total charge $e$ of the bosons). However, if we start with 
$[11]$, then a domain wall with either $[02]$ or $[20]$ can localize the charge $e/2$ quasihole. 
Therefore, the adjacency matrix for this quasihole is given by:
\begin{eqnarray}
 A=\para{\begin{array}{ccc}
     0 & 0 & 1  \\ 
     0 & 0 & 1 \\
     1 & 1 & 0 \\ 
  \end{array}},
\end{eqnarray}
where the rows/columns refer to $[20]$, $[02]$, and $[11]$, respectively. 

The quantum dimension of this quasihole corresponds to the largest eigenvalue of $A$, which is $\sqrt{2}$. 
We see that $A$ corresponds exactly the fusion rules of the non-Abelian $\sigma$ quasiparticle in the bosonic Moore-Read Pfaffian
FQH state. 

Furthermore, we observe that there are topologically non-trivial charge $e$ excitations, associated with domain walls between
$[20]$ and $[02]$ or vice versa, and between $[11]$ and itself. Physically, these corresopnd to inserting the Laughlin $e/2$ quasiholes
in both layers simultaneously. In the $(220)$ state, the excitation associated with inserting charge $e/2$ in each layer has fermionic
statistics. Since this excitation is unaffected by the phase transition because it is $Z_2$ layer symmetric, it continues to be a fermion
after the transition. 

Therefore, from the thin torus limit we have found that the non-Abelian phase consists of three topological classes of excitations:
the local ones, an Abelian fermion, and a non-Abelian charge $e/2$ quasihole with quantum dimension $2$.
This agrees with the topological order of both the Moore-Read Pfaffian FQH state, and the $SU(2)_2$ parton CS theory
described by the $U(4)_1/SU(2)_2$ edge theory. 

The above scheme can be easily explored for general $(nn0)$ case. The $(nn0)$ state has a ground state degeneracy of $n^2$. 
After the phase transition driven by the interlayer tunneling, we keep only the $n(n+1)/2$ $Z_2$ layer symmetric combinations. 
The adjacency matrix for the quasihole with $q=e/n$ can be easily found. We find that the largest eigenvalue, which sets the
quantum dimension of the quasihole operator, is $d_{qh}=2\cos\para{\frac{\pi}{n+2}}$. This is exactly the quantum dimension 
of the most relevant primary field in the $SU(n)_{2}$ chiral WZW model! More generally, we find that the adjacency 
matrices of the $n(n+1)/2$ types of excitations coincide with the fusion rules
of the representation algebra of $SU(n)_2$. 

{\subsection{b. Adjacency matrix for all excitations in the Fibonacci phase}

In this section we consider the non-Abelian phase of $2/3$ bilayer state with interlayer tunneling and compute the adjacency matrix associated with all nontrivial excitations. In this phase, there are six degenerate ground-states on the torus geometry with the following thin torus patters: $[200],[020],[002],[110],[101],$ and $[011]$, where $[200]\equiv \left[\begin{array}{c}
     100  \\ 
     100 \\ 
  \end{array}\right]$, $[110]\equiv \left[\para{\begin{array}{c}
     100  \\ 
     010 \\ 
  \end{array}}+\para{\begin{array}{c}
     010  \\ 
     100 \\ 
  \end{array}}\right]$, and so on. Recall that the quasiparticles can be understood as domain walls between the different ground state patterns. 
If we start with the state $[200]$ and consider a domain wall with the state $[110]$, then from the Su-Schrieffer
counting argument we see that there is a charge $e/3$ quasihole. This can be understood as the original 
Laughlin $e/3$ quasihole, but inserted in either the top layer or the bottom layer, with equal weight. 
If instead we start with the state $[110]$ and consider a domain wall with either $[020]$ or $[101]$, 
we see that there is again a charge $e/3$ quasihole. In general, we can ask which pairs of ground states, 
labelled $i$ and $j$, give rise to a charge $e/3$ quasihole at their domain wall. This defines an 
adjacency matrix for the charge $e/3$ quasihole,
\begin{eqnarray}\label{Eq:Adj-1}
 &&A_{e/3}=\para{\begin{array}{cccccc}
 0&0  &0 &1&0&0 \\
 0&0  &0 &0&0&1 \\
 0&0  &0 &0&1&0 \\
 0&1  &0 &0&1&0 \\
 1&0  &0 &0&0&1 \\
 0&0  &1 &1&0&0 \\
  \end{array}}
 \end{eqnarray}
where the rows/columns correspond to $[200],[020],[002],[110],[101],[011]$, respectively. 
More generally, let us consider $n_{e/3}$ quasi-holes with $q=e/3$ at positions
$j_1,j_2,\cdots, j_{n_{e/3}}$ \cite{ardonne2008}. To do so, we start with a fixed ground-state pattern, 
say $[200]$. At site $j_1$, there is a domain wall with $[110]$, at site $j_2$ 
there can be either $[020]$ or  $[101]$ patterns, and so on. We see that the number 
of possibilities grows exponentially with $n_{e/3}$. It is straightforward 
to verify that ${\bf tr}\para{A^{n_{e/3}}}$ gives the total number of different possibilities on the torus. 
Therefore, the degeneracy of the ground-state in the presence of $n_{e/3}$ quasihole insertions 
grows as $\lambda_1^{n_{e/3}}$ where $\lambda_{1}$ is the dominant eigenvalue of the adjacency 
matrix $A$. Consequently, the quantum dimension of the quasihole operator with minimum 
electric charge is $\lambda_1$. Using the above adjacency matrix, the quantum dimension of 
the charge $e/3$ quasihole is the golden ratio: $d_{e/3}=F\equiv \frac{1+\sqrt{5}}{2}$. 

Now consider the $-e/3$ charge excitation where inserts $-e/3$ charge on the top or bottom layers with equal weights. From the Su-Schrieffer counting it is straightforward to check that the domain wall between $[200]$ and $[101]$ states carries the desired charge. However, $[101]$ state can be followed by either $[200]$ or $[110]$ states. The full adjacency matrix for this excitation can be obtained after which we have:

\begin{eqnarray}\label{Eq:Adj-2}
 &&A_{-e/3}=\para{\begin{array}{cccccc}
 0&0  &0 &0&1&0 \\
 0&0  &0 &1&0&0 \\
 0&0  &0 &0&0&1 \\
 1&0  &0 &0&0&1 \\
 0&0  &1 &1&0&0 \\
 0&1  &0 &0&1&0 \\
  \end{array}}
\end{eqnarray}
Again, the dominant eigenvalue of the above matrix is the golden ratio, so the corresponding operator is non-Abelian with $d_{-e/3}=F$ quantum dimension.

Now let us consider an overal neutral excitation where inserts either $e/3$ ($-e/3$) charge on the top (bottom) layer or $-e/3$ ($e/3$) charge on the top (bottom) layer with equal weights. From the Su-Schrieffer counting it is easy to check that the domain wall between $[200]$ and $[011]$ states has vanishing total electric charge. However, starting with $[011]$ we have two possibilities: we can put either $[011]$ state itself or $[200]$ state next to it. Similarly, we can find the full adjacency matrix which would be: 
\begin{eqnarray}\label{Eq:Adj-3}
 &&A_{q=0}=\para{\begin{array}{cccccc}
 0&0  &0 &0&0&1 \\
 0&0  &0 &0&1&0 \\
 0&0  &0 &1&0&0 \\
 0&0  &1 &1&0&0 \\
 0&1  &0 &0&1&0 \\
 1&0  &0 &0&0&1 \\
  \end{array}}
 \end{eqnarray}
 Again, the dominant eigenvalue of the above matrix is the golden ratio, so the corresponding operator is non-Abelian with $d_{q=0}=F$ quantum dimension.

Now, let us consider the operator which inserts two charge $e/3$ excitations each on one layer. The adjacency matrix associated with this excitation is:
\begin{eqnarray}\label{Eq:Adj-4}
 &&A_{2e/3}=\para{\begin{array}{cccccc}
 0&1  &0 &0&0&0 \\
 0&0  &1 &0&0&0 \\
 1&0  &0 &0&0&0 \\
 0&0  &0 &0&0&1 \\
 0&0  &0 &1&0&0 \\
 0&0  &0 &0&1&0 \\
  \end{array}}
 \end{eqnarray}
whose eigenvalues are all unimodular. So the quantum dimension of these excitations is $d_{2e/3}=1$, hence they are Abelian excitations. 

Finally, let us consider the charge $4e/3$ excitation which creates two $2e/3$ excitations each on one layer. The corresponding adjacency matrix is:
\begin{eqnarray}\label{Eq:Adj-5}
 &&A_{4e/3}=\para{\begin{array}{cccccc}
 0&0  &1 &0&0&0 \\
 1&0  &0 &0&0&0 \\
 0&1  &0 &0&0&0 \\
 0&0  &0 &0&1&0 \\
 0&0  &0 &0&0&1 \\
 0&0  &0 &1&0&0 \\
  \end{array}}
 \end{eqnarray}
whose eigenvalues are all unimodular again. So the quantum dimension of these excitations is $d_{2e/3}=1$ as well. Thus, the corresponding operators are Abelian. 
}

\subsection{c. Generalization to $k$ layers}

Let us consider the edge theory of $k$ copies of the $1/n$ Laughlin FQH state. This includes $k$ chiral bosons, 
$\varphi_i$, $i = 1,...,k$, with the local `electron' operator in each layer given by
$\Psi_{ei} = e^{i n \varphi_i}$. There are $n^k$ topologically distinct quasiparticles, labelled by the vertex operators
\begin{align}
V_{\vec{a}} = e^{i a_i \varphi_i } ,
\end{align}
where $\vec{a}$ is a $k$-component vector, with each entry $a_i = 0, ..., n-1$. 

Next, let us consider symmetrizing these vertex operators:
\begin{align}
\Phi_{\vec{a}} = \sum_{P} V_{P(\vec{a})},
\end{align}
where $\sum_P$ is the sum over permutations of $k$ layers, where $P(\vec{a})$ is the permuted vector. Due to the permutation, we obtain only $\para{  \begin{array}{c}
    n+k-1 \\ 
    k \\ 
  \end{array}}$ distinct quasi-particle operators. Remarkably, we find that the fusion rules $\Phi_{\vec{a}} \times \Phi_{\vec{a}'}$ coincide with the fusion rules
of the representation algebra of the quantum group $SU(n)_k$! This is closely related to the observation of
\cite{ardonne2009}, where it was found that certain types of occupation patterns give rise to adjacency matrices
that are related to the fusion rules of $SU(n)_k$. This result is a straightforward application of the methods presented in this paper. 

\subsection{d. Generalization to $(nnl)$ bilayer state}
For the $(nnl)$ Halperin state we can define the charged and neutral chiral bosons as $\phi_{c}=\frac{\phi_{1}+\phi_{2}}{\sqrt{2}}$, and $\phi_{s}=\frac{\phi_{1}-\phi_{2}}{\sqrt{2}}$. Using the $K$ matrix associated with the $(nnl)$ state, the `chiral' charged (neutral) boson is compactified on a circle with $R_{c}=\sqrt{\frac{n+l}{2}}$ ($R_{s}=\sqrt{\frac{n-l}{2}}$) radius. Using the methods developed in this paper, we can study the effect of uniform interlayer tunneling and show that the quasiparticles follow $SU(n-l)_2$ fusion algebra for strong enough tunneling. Accordingly, the fusion rule together with the level-rank duality imply that there is a phase transition to $U(1)_{2\para{n+l}} \times SU(2)_{n-l}$ non-Abelian state.  

\subsection{e. Duality between interlayer tunneling and interlayer pairing}
Let us consider an $(nnl)$ Halperin state with interlayer pairing. The electron operators in this state are $\Psi_1=e^{i\para{R_c\phi_c+R_s\phi_s}}$ and $\Psi_2=e^{i\para{R_c\phi_c-R_s\phi_s}}$, where $\phi_{c/s}=\frac{\phi_1\pm \phi_2}{\sqrt{2}}$, and $R_{c/s}=\sqrt{\frac{n\pm l}{2}}$. Next, we perform a particle-hole transformation on the bottom layer after which $\phi_{1} \to \phi_1$, $\phi_2 \to -\phi_2$, thus $\phi_c \leftrightarrow \phi_s$. Doing so, we obtain an $(nn,-l)$ bilayer state whose electron operators are $\overline{\Psi}_1=e^{i\para{\overline{R}_c\overline{\phi}_c+\overline{R}_s\overline{\phi}_s}}$ and $\overline{\Psi}_2=e^{i\para{\overline{R}_c\overline{\phi}_c-\overline{R}_s\overline{\phi}_s}}$, where $\overline{\phi}_{c/s}=\phi_{s/c}$, and $\overline{R}_{c/s}=R_{s/c}$. Therefore, electron operators transform as $\Psi_1 \to \overline{\Psi}_1$ and $\Psi_2 \to \overline{\Psi}_2^\dag$. The benefit of the particle-hole transformation is that interlayer pairing becomes interlayer tunneling in the transformed state. For example, $\Psi_{2,R} \Psi_{1,L}\to \overline{\Psi}_{2,R}^\dag \overline{\Psi}_{1,L}$. Therefore, the topological order of the $(nnl)$ bilayer state with interlayer pairing is expected to be equivalent to that of the $(n,n,-l)$ state with strong interlayer pairing.

\end{widetext}


\end{document}